\documentclass[10pt]{article}

\usepackage{amsmath}
\usepackage{amssymb}

\usepackage{graphicx}

\usepackage{cite}

\usepackage{color} 

\usepackage[labelfont=bf,labelsep=period,justification=raggedright]{caption}

\makeatletter
\renewcommand{\@biblabel}[1]{\quad#1.}
\makeatother

\date{}

\newcommand{\toyLIFE}{{\tt \fontfamily{qcr}\selectfont t\!\raisebox{-.1em}{\fontfamily{qcr}\selectfont o}\!\fontfamily{qcr}\selectfont y}\textsf{\fontfamily{phv}\selectfont LIFE}}

\begin{document}

\begin{flushleft}
{\Large\toyLIFE
\textbf{: a computational framework to study the multi-level organization of the genotype-phenotype map}
}
\\
Clemente F. Arias$^{1,2}$, 
Pablo Catal\'an$^{1,2}$, 
Susanna Manrubia$^{1,3}$, 
Jos\'e A. Cuesta$^{1,2,4\ast}$
\\
\bf{1} Grupo Interdisciplinar de Sistemas Complejos (GISC), Madrid, Spain
\\
\bf{2} Dept. Matem\'aticas, Universidad Carlos III de Madrid,
Legan\'es, Madrid, Spain
\\
\bf{3} Centro Nacional de Biotecnolog\'{\i}a (CSIC), Campus de 
Cantoblanco, Madrid, Spain
\\
\bf{4} Instituto de Biocomputaci\'on y F\'{\i}sica de Sistemas Complejos
(BIFI), Universidad de Zaragoza, Zaragoza, Spain
\\
$\ast$ E-mail: Corresponding cuesta@math.uc3m.es
\end{flushleft}

\section*{Abstract}

The genotype-phenotype map is an essential object in our understanding of
organismal complexity and adaptive properties, determining at once genomic
plasticity and those constraints that may limit the ability of genomes to
attain evolutionary innovations. An exhaustive experimental characterization of
the relationship between genotypes and phenotypes is at present out of reach.
Therefore, several models mimicking that map have been proposed and
investigated, leading to the identification of a number of general features:
genotypes differ in their robustness to mutations, phenotypes are represented
by a broadly varying number of genotypes, and simple point mutations seem to
suffice to navigate the space of genotypes while maintaining a phenotype.
However, most current models address only one level of the map (sequences and
folded structures in RNA or proteins; networks of genes and their dynamical
attractors; sets of chemical reactions and their ability to undergo molecular
catalysis), such that many relevant questions cannot be addressed. Here we
introduce \toyLIFE{}, a multi-level model for the genotype-phenotype map based
on simple genomes and interaction rules from which a complex behavior at upper
levels emerges, remarkably plastic gene regulatory networks and metabolism.
\toyLIFE{} is a tool that permits the investigation of how different levels are
coupled, in particular how and where do mutations affect phenotype or how the
presence of certain metabolites determines the dynamics of \toyLIFE{} gene
regulatory networks. The possibilities of this model are not exhausted by the
results presented in this contribution. It can be easily generalized to
incorporate evolution through mutations that change genome length or through
recombination, to consider gene duplication or deletion, and therefore to
explore further properties of extended genotype-phenotype maps.

\section*{Author Summary}

How the measurable traits of organisms, their phenotypes, emerge from the
instructions encoded in genomes, counts amongst the most challenging and
essential questions to understand organismal evolution and adaptation. At
present, the genotype-phenotype relationship cannot be experimentally evaluated
in an exhaustive fashion, since research is mostly limited to analyse the
effect that specific mutations have in the fitness of organisms. Therefore,
most of our knowledge has been derived from simple models that investigate
one-level maps, as RNA sequences and their folded states, or ensembles of
chemical reactions and the molecules they can metabolize. Given our current
limitations, it is essential to work with models that are as realistic as
possible, but keep computational tractability. Here we introduce \toyLIFE{}, a
multi-level model that maps genotypes (formed by a variable number of genes and
a polymerase protein) into phenotypes. Once interactions between proteins in
the model and environmental metabolites are defined, phenotypes correspond to
those ensembles of genes able to undertake metabolite catalysis. Other natural
properties such as regulatory networks or proteins with multiple functions
naturally emerge in \toyLIFE{}, which appears as a powerful tool to explore
aspects of the genotype-phenotype map that as yet cannot be experimentally
addressed.

\section*{Introduction}

Describing and understanding the intricacies of the genotype-phenotype map
counts amongst the most difficult and most essential issues to comprehend
organismal complexity and adaptation through natural selection
\cite{lewis:2012}. High-throughput data obtained from whole genome sequencing
and other -omics techniques currently allow a characterization with
unprecedented detail of how genotypic variation affects phenotypes. The
analysis of gene networks has demonstrated that phenotypes cannot be understood
on the basis of isolated genes \cite{karlebach:2008}, and that the effects of
mutations strongly depend on a genetic background that expresses at different
levels before generating a final phenotype \cite{chandler:2013}. While simple
point mutations may affect more than one gene \cite{gwagner:2011}, phenotypes,
overall, tend to be extremely robust: populations may sustain a high level of
criptic variation that acts at once as a buffering mechanism
\cite{rutherford:2000} and as a reservoir of variability to promote rapid
adaptation \cite{paaby:2014}. An essential part of our improved understanding
of the concepts, design principles and general mechanisms underlying the
appearance of biological function from organismal genomes arises from the use
of {\it in silico} tools and models \cite{diVentura:2006}. 
 
The neutral theory of evolution \cite{kimura:1968,kimura:1984} posits that most
mutations have no, or very little, effect on phenotypes, and are thus ignored
by natural selection. This is an amply supported fact, though the level at
which mutations cease to have an effect is a matter of research. DNA is
translated into proteins and ribozymes which fold into three-dimensional
structures. These molecules bind to each other and to the genome itself,
enhancing or inhibiting the expression of genes ---hence forming highly complex
regulatory networks---, and eventually interact with metabolites to produce the
metabolic pathways that sustain cellular life \cite{watson:2013}. Redundancy
appears at all these levels. Besides the well-known redundancy of the genetic
code, many different aminoacid \cite{lipman:1991} ---in the case of proteins---
or RNA  \cite{schuster:1994} ---in the case of ribozymes--- sequences fold into
equivalent three-dimensional structures and exhibit similar interaction sites,
thus maintaining their functions. Regulatory and metabolic networks are quite
robust to additions, eliminations or substitutions of some of their components
as well. For instance, regulatory regions with similar transcriptional output
often have little overt sequence similarity, both within and between genomes
\cite{weirauch:2010}. Also, regulatory DNA sequences in different
\emph{Drosophila} species exhibiting the same expression patterns are not
conserved  \cite{hare:2008}. As of robustness of metabolic networks, one-gene
knockout experiments with \emph{Saccharomyces cerevisiae} show that around 50\%
of mutants show a selective disadvantage below 1\% relative to the wildtype
\cite{thatcher:1998}. Similar results have been obtained with 
\emph{Escherichia coli}  \cite{baba:2006}.

The huge number of genomic solutions ushering in the same phenotype leads to
the concept of genotype networks, that is ensembles of genotypes that yield the
same phenotype and can be mutually accessed through mutations
\cite{wagner:2011b}. Genotype networks often traverse the whole space of
genotypes, and are highly interwoven: virtually any phenotype is just a few
mutations away from any other. These networks reflect the robustness of
phenotypes against mutations, and their structure is essential to promote
adaptability and evolutionary innovation \cite{draghi:2010,wagner:2011}. Most
of our knowledge on the topology of genotype networks relies on information
obtained from well-motivated computational models that map genotype onto a
simplified phenotype. Classical examples mapping sequence to molecular
structure (which acts as a proxy for phenotype) are those of RNA
\cite{schuster:2006} or proteins folded through algorithms of variable
complexity \cite{dill:1985,bastolla:2000}. Other models have addressed the map
between higher expression levels, as those mimicking gene regulatory networks
\cite{kauffman:1993,ciliberti:2007,payne:2013} or metabolism
\cite{matias:2009}. 

Despite the significant conceptual advances provided by those models, there are
two crucial elements of the genotype-phenotype map that they disregard: the
existence of a hierarchy of expression levels between genotype and phenotype
and the bi-directional coupling among the levels. Studies focusing on RNA or
proteins assume that the molecular function is mostly determined by their
spatial structure. This makes sense for some very specific enzymes
\cite{schultes:2000} but, in general, these molecules are pieces of complex
regulatory or metabolic networks. Further, molecular interactions are not
considered (see \cite{greenbury:2014} for an exception modelling the quaternary
structure of proteins), and there is no representation of the molecular context
\cite{piatigorsky:2007}. Therefore, cases where a protein may act as an
enhancer of the expression of a gene by sticking to its promoter, but may
become sequestered and thus inactivated in the presence of another protein are
impossible to embody in one-molecule models, among many others.

In turn, models considering higher levels typically disregard the dynamics of
underlying sequences. Gene regulatory networks are represented in an effective
way through direct interactions among their components, as in Boolean
regulatory networks. In this case, gene states are binary variables which
interact (enhancing or inhibiting the expression of interacting genes) to
determine new states at a subsequent time step \cite{kauffman:1969,cheng:2011}.
Boolean networks do not consider how mutations at the genome level propagate to
upper levels, and only implement straight changes in the Boolean functions. The
situation is similar with metabolic models that use the ensemble of metabolic
reactions as genotypes, since the kind of mutations considered can thus only be
the elimination or addition of reactives or full reactions \cite{matias:2009}. 

The current situation is that we lack a model that captures the essentials of
the biology at all levels from genome to metabolisms, but which at the same
time is sufficiently simple so as to provide useful answers and insights about
the genotype-phenotype mapping. In this paper we make one such proposal, that
we refer to as \toyLIFE. \toyLIFE{} is a model that contains simplified
versions of genes, promoters, proteins, and metabolites, which interact with
each other under the laws of a simplified chemistry. Besides introducing the
model and showing examples of its rich phenomenology, we identify a number of
emerging properties that \toyLIFE{} shares with natural systems. Such are the
existence of a large number of robust phenotypes, of common metabolic functions
(which arise in the absence of any evolutionary fine-tuning), a space-covering
map at the sequence-structure level (as observed in RNA and protein folding
models) but a small fraction of metabolically functional genomes. Coupling
among different levels restricts the diversity of possible Boolean functions,
as well as the metabolites that can be broken. In the framework of \toyLIFE,
mutations can show their effects at different levels before affecting the
phenotype, and functional molecules can be co-opted to fulfill different and
not previously foreseen functions. 

\section*{Results}

\subsection*{Definition of \toyLIFE{}}

The basic building blocks of \toyLIFE{} are toyNucleotides (toyN),
toyAminoacids (toyA), and toySugars (toyS). Each block comes in two flavors:
hydrophobic (H) or polar (P). Random polymers of basic blocks constitute
toyGenes (formed by 20 toyN units), toyProteins (chains of 16 toyA units), and
toyMetabolites (sequences of toyS units of arbitrary length). These elements of
\toyLIFE{} are defined on the two-dimensional space (Figure
\ref{fig:elements}).

\subsubsection*{toyGenes}
toyGenes are composed of a 4-toyN promoter region followed by a 16-toyN coding
region. There are $2^4$ different promoters and $2^{16}$ coding regions,
leading to $2^{20}\approx 10^6$ toyGenes. An ensemble of toyGenes forms a
genotype. If the toyGene is expressed, it will produce a chain of 16 toyA that
represents a protein.  Translation follows a straightforward rule: H (P) toyN
translate into H (P) toyA. 

\subsubsection*{toyProteins}
\begin{figure}[!b] 
\centerline{\includegraphics[width=80 mm]{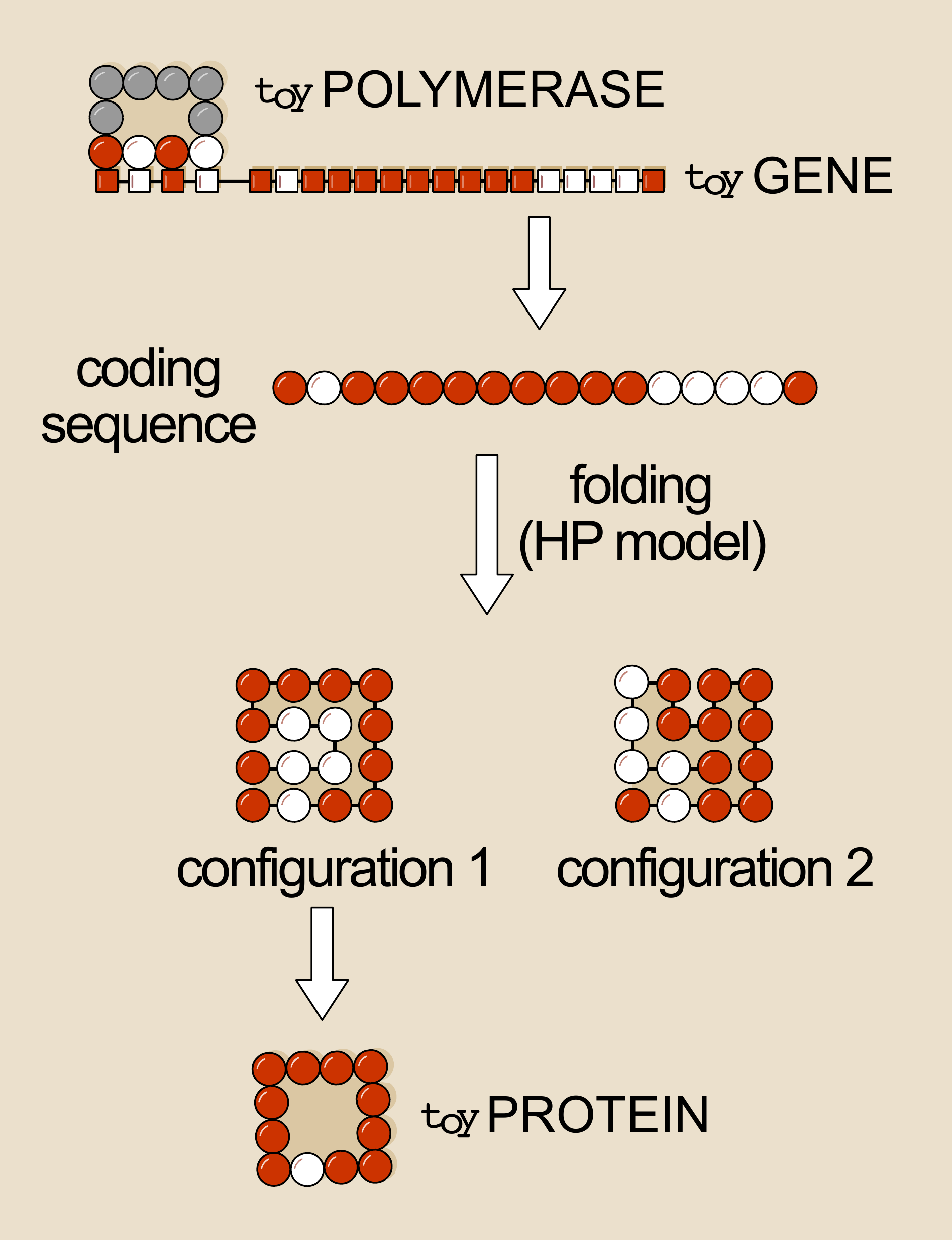}}
\caption[]{\small\textbf{Folding of a toyProtein.} When a toyGene is expressed, its coding region is translated into a sequence of toyAminoacids, which folds on 
a $4\times 4$ lattice following a self-avoiding walk. As a result, the toyProtein acquires a folding energy, which is the sum of the interaction energies between
non-contiguous toyAminoacids of the chain (one, two or three, with
energies ranging from 0 to $-2$). Interaction energy is pairwise additive. A toyProtein folds into the 
structure that minimizes this folding energy. If two structures have the same minimal folding energy, the one with the minimum number of H toyAminoacids on its
perimeter is chosen; if this number also coincides, the toyProtein does not fold. toyProteins are therefore characterized by two traits: their perimeter and 
their folding energy.}
\label{fig:folding}
\end{figure}

toyProteins correspond to the minimum energy, maximally compact folded
structure of the 16 toyA chain arising from a translated gene. Their folded
configuration is calculated through the hydrophobic-polar (HP) protein lattice
model \cite{dill:1985} (see Methods and Figure \ref{fig:folding}). Two toyProteins can bind to
each other to form a toyDimer, which is the only protein aggregate considered
in \toyLIFE{}. The toyPolymerase is a special kind of molecule characterized by
a specific sequence of 4 toyA and an otherwise indetermined structure. It is
always present in the system. In the \toyLIFE{} universe, only the folding
energy and perimeter of a toyProtein matter to characterize its interactions,
so folded chains sharing these two features are indistinguishable. This is a
difference with respect to the original HP model, where different inner cores
defined different proteins and the composition of the perimeter was not
considered as a phenotypic feature. However, subsequent versions of HP had
already included additional traits \cite{hoque:2009}.

\subsubsection*{Interactions and dynamics}
\begin{figure}[!t] 
\centerline{\includegraphics[width=165 mm]{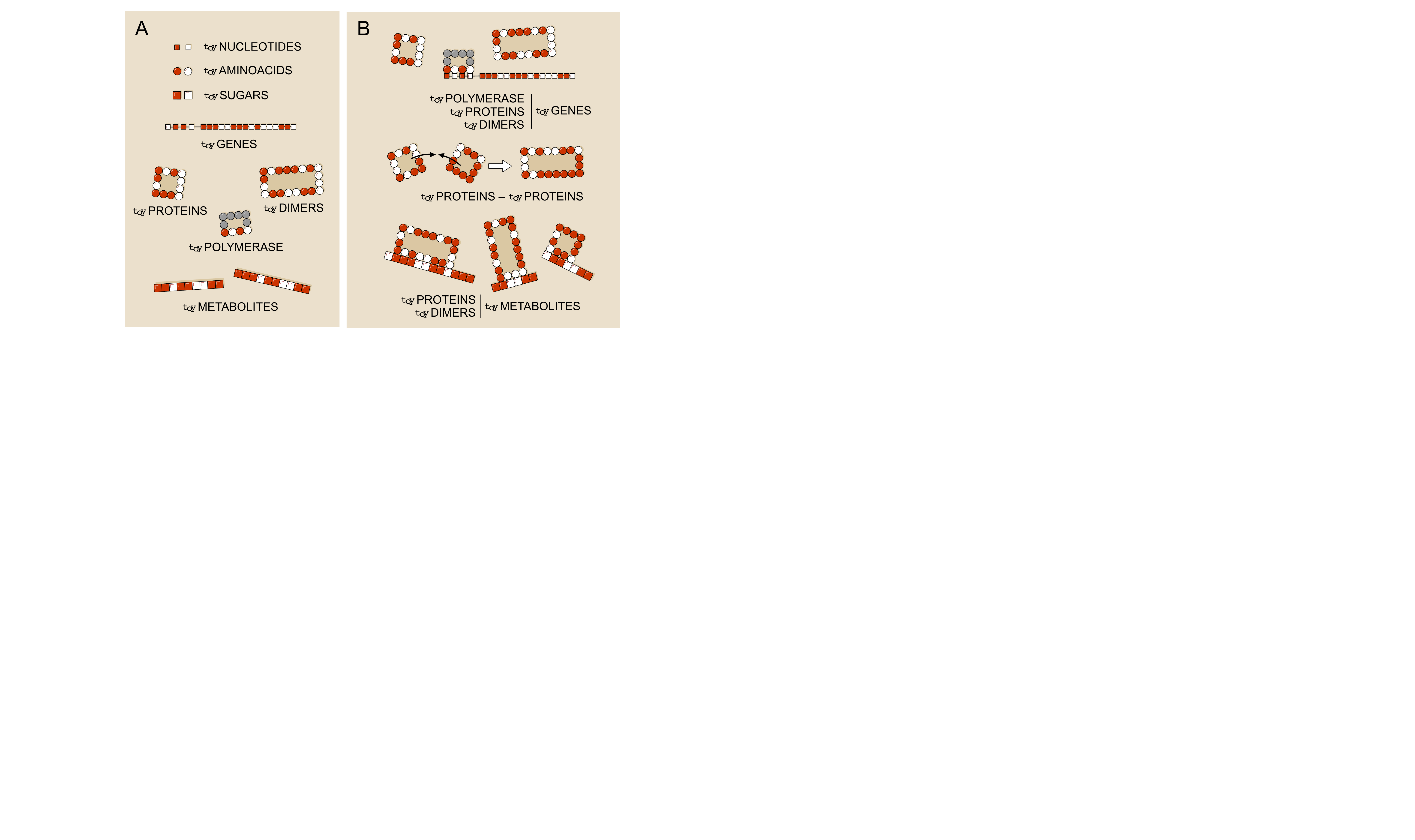}}
\caption[]{\small\textbf{Building blocks and interactions defining \toyLIFE.} \textbf{A:} The three basic building blocks of \toyLIFE{} are toyNucleotides,
toyAminoacids, and toySugars. They can be hydrophobic (H, white) or polar (P, red), and their random polymers constitute toyGenes, toyProteins, and toyMetabolites.
\textbf{B:} Possible interactions between pairs of \toyLIFE{} elements. toyGenes interact through their promoter region with toyProteins (including the 
toyPolymerase and toyDimers); toyProteins can bind to form toyDimers, and interact with the toyPolymerase when bound to a promoter; both toyProteins and toyDimers 
can bind a toyMetabolite at arbitrary regions along its sequence.}
\label{fig:elements}
\end{figure}

Expression of toyGenes occurs through the interaction with the toyPolymerase,
which can bind to promoters but also to other toyProteins (including toyDimers,
see Methods and Figure \ref{fig:elements}\textbf{A}). When the toyPolymerase
binds to a promoter, translation is directly activated and the corresponding
toyGene is expressed. However, a more stable (lower energy) binding of a
toyProtein or toyDimer to a promoter precludes the binding of the
toyPolymerase. This inhibits the expresion of the toyGene, except if the
toyPolymerase binds to a specific site of the toyProtein/toyDimer, in which
case the toyGene can be expressed (Figure \ref{fig:regulation}). 

Following the rules of the HP model, H-P and H-H interactions are also used for
inter-molecular interactions, and thus to quantify the energy of the bound
states toyProtein-toyGene, toyProtein-toyProtein, and toyProtein-toyMetabolite
(see Methods and Figure \ref{fig:elements}). Both the version of the HP model
we use and the inter-molecular interactions considered depend on parameters
that stand for the decrease in free energy when H-P and H-H bonds are formed:
$E_{HP}=-2$ and $E_{HH}=-0.3$, respectively (with $E_{PP}=0$), as in
\cite{li:1996}.  

\begin{figure}[!t]
\centerline{\includegraphics[width=150mm]{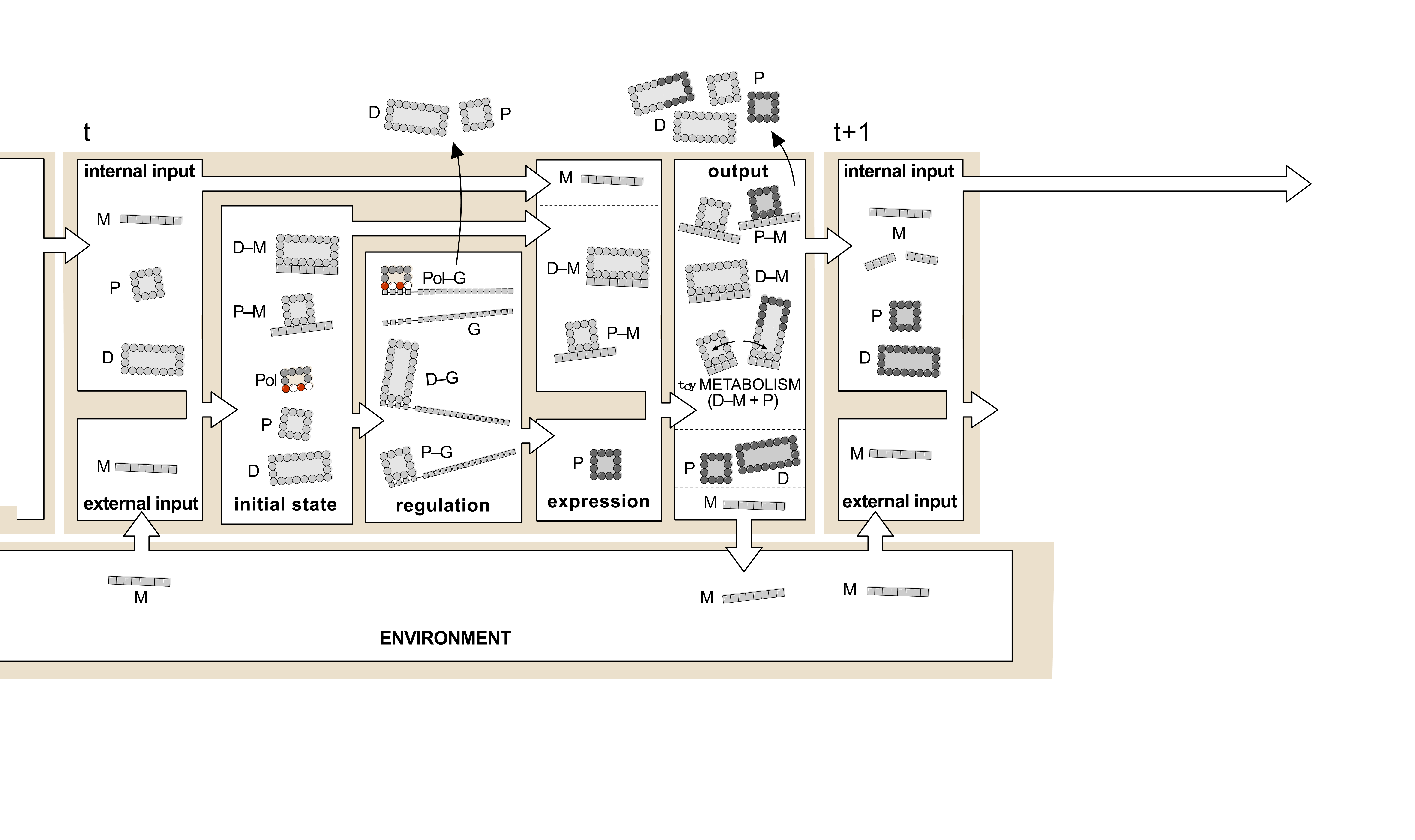}}
\caption[]{\small\textbf{Dynamics of \toyLIFE.} Input molecules at time step $t$ are toyProteins (Ps) (including toyDimers (Ds)) and toyMetabolites, either produced 
as output at time step $t-1$ or environmentally supplied (all toyMetabolites denoted Ms). Ps and Ds interact with Ms to produce complexes P-M and D-M. Next, these 
complexes, the remaining Ps and Ds, and the toyPolymerase (Pol) interact with toyGenes (G) at the regulation phase. The most stable complexes with promoters are 
formed (Pol-G, P-G and D-G), activating or inhibiting toyGenes. P-Ms and D-Ms do not participate in regulation. Ps and Ds not in complexes are eliminated and 
new Ps (dark grey) are formed. These Ps interact with all molecules present and form Ds, new P-M and D-M complexes, and catalize old D-M complexes. At the end of 
this phase, all Ms not bound to Ps or Ds disappear, and all Ps and Ds in P-M and D-M complexes unbind and also disappear. The remaining molecules (Ms just released
from complexes, as well as all free Ps and Ds) go to the input set of time step $t+1$.}
\label{fig:iteration}
\end{figure}

The interaction and dynamical rules in \toyLIFE{} have been chosen so as to
make the model as simple as possible, while retaining the essentials of
molecular genetics. Following the original HP model (which disregards lattice
proteins with multiple minimal energy folds), interactions between \toyLIFE{}
elements are univocally defined according to specific disambiguation rules (see
Methods). The dynamics of the model proceeds in discrete time steps and
variable molecular concentrations are not taken into account. A step-by-step
description of \toyLIFE{} dynamics is summarized in Fig. \ref{fig:iteration}.
There is an initial set of molecules which results from the previous time step:
toyProteins (including toyDimers and the toyPolymerase) and toyMetabolites,
either endogenous or provided by the environment. These molecules first
interact between them to form possible complexes (see Methods) and are then
presented to a collection of toyGenes that is kept constant along subsequent
iterations. Regulation takes place, mediated by a competition for binding the
promoters of toyGenes, possibly causing their activation and leading to the
formation of new proteins.  toyProteins/toyDimers not bound to any
toyMetabolite disappear in this phase. Thus, only the newly expressed
toyProteins and the complexes involving toyMetabolites in the input set remain.
All these molecules interact yet again (see Methods), and here is where
catalysis can occur (see below). Free toyProteins/toyDimers will form part of
the input set for the next time step. However, toyProteins/toyDimers bound to
toyMetabolites disappear in this phase, and only the toyMetabolites are kept as
input to the next time step. Unbound toyMetabolites are returned to the
environment.

\subsection*{toyProteins behave as toyGene switches}

\begin{figure}[!t] 
\centerline{\includegraphics[angle=270,width=150 mm]{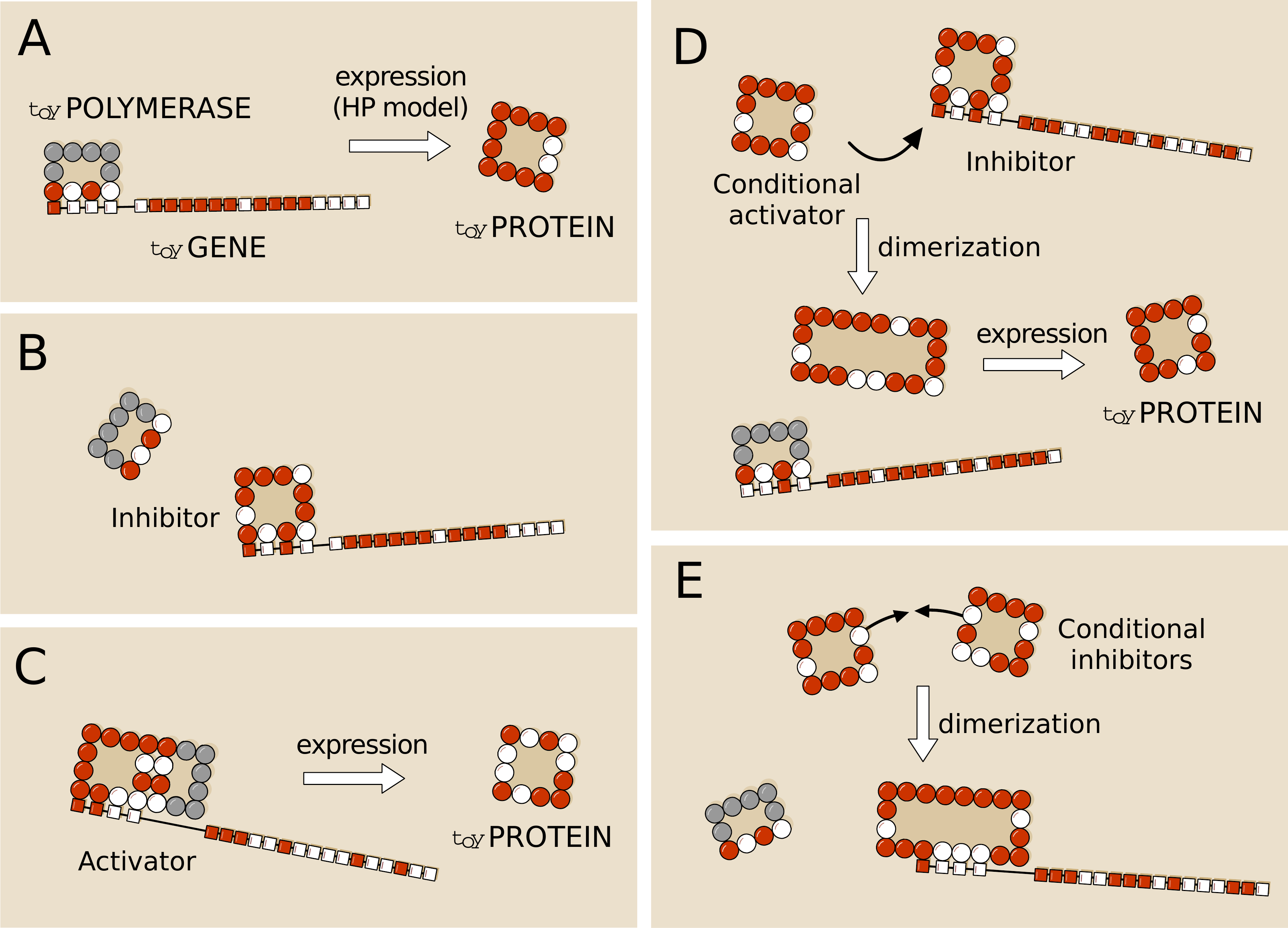}}
\caption[]{\small\textbf{Emergent functions in toyProteins.} \textbf{A:} A toyGene is expressed (translated) when the toyPolymerase binds to its promoter region.
The sequence of Ps and Hs of the toyProtein will be exactly the same as that of the toyGene coding region. \textbf{B:} If a toyProtein binds to the promoter 
region of a toyGene with a lower energy than the toyPolymerase does, it will displace the latter, and the toyGene will not be expressed. This toyProtein acts as 
an \emph{inhibitor}. \textbf{C:} The toyPolymerase does not bind to every promoter region. Thus, not all toyGenes are expressed constitutively. However, some
toyProteins will be able to bind to these promoter regions. If, once bound to the promoter, they bind to the toyPolymerase with their rightmost side, the
toyGene will be expressed, and these toyProteins act as
\emph{activators}. \textbf{D:} More complex interactions ---involving
more elements--- appear. For
example, a toyProtein that forms a toyDimer with an inhibitor
---preventing it from binding to the promoter--- will effectively activate the expression of the
toyGene. However, it does neither interact with the promoter region nor with the toyPolymerase, and its function is carried out only when the inhibitor is
present. We call this kind of toyProteins \emph{conditional activators}. \textbf{E:} Finally, two toyProteins can bind together to form a toyDimer that
inhibits the expression of a certain toyGene. As they need each other to perform this function, we call them \emph{conditional inhibitors}. As the number of
genes increases, this kind of complex relationships can become very intricate.}
\label{fig:regulation}
\end{figure}

The minimal interaction rules that define \toyLIFE{} dynamics endow toyProteins
with a set of possible activities not included {\it a priori} in the rules of
the model (see Figure \ref{fig:regulation}). For example, since the 4-toyN interacting site of
the toyPolymerase cannot bind to all promoter regions, translation mediated by
a toyProtein or toyDimer binding might allow the expression of genes that would
otherwise never be translated. These toyProteins thus act as activators. This
process finds a counterpart in toyProteins that bind to promoter regions more
stably than the toyPolymerase does, and therefore prevent gene expression. They
are acting as inhibitors. There are two additional functions that could not be
foreseen and involve a larger number of molecules. A toyProtein that forms a
toyDimer with an inhibitor ---preventing its binding to the promoter---
effectively behaves as an activator for the expression of the toyGene. However,
it interacts neither with the promoter region nor with the toyPolymerase, and
its activating function only shows up when the inhibitor is present. This kind
of toyProteins thus act as conditional activators. On the other hand, two
toyProteins can bind together to form a toyDimer that inhibits the expression
of a particular toyGene.  As the presence of both toyProteins is needed to
perform this function, they behave as conditional inhibitors. This flexible,
context-dependent behavior of toyProteins, permits the construction of toy Gene
Regulatory Networks (toyGRN). 

\subsection*{Gene regulatory networks in \toyLIFE{} are deterministic Boolean
networks}

Molecular interactions and dynamical rules in \toyLIFE{} can be translated into
toyGRN that behave as deterministic Boolean networks
\cite{kauffman:1969,cheng:2011}. The corresponding Boolean variables are the
states (expressed or not expressed) of toyGenes. These variables are
transformed through Boolean functions that represent the dynamical rules
described, having as input current toyGene states and as output their states at
the next time step. Boolean functions depend on the toyProteins present in the
system and on the functions they perform. Through iteration of the Boolean map
one can characterize the set of attractors of the dynamics and the
corresponding basins of attraction.

\begin{figure}[!t]
\centerline{\includegraphics[width=150mm]{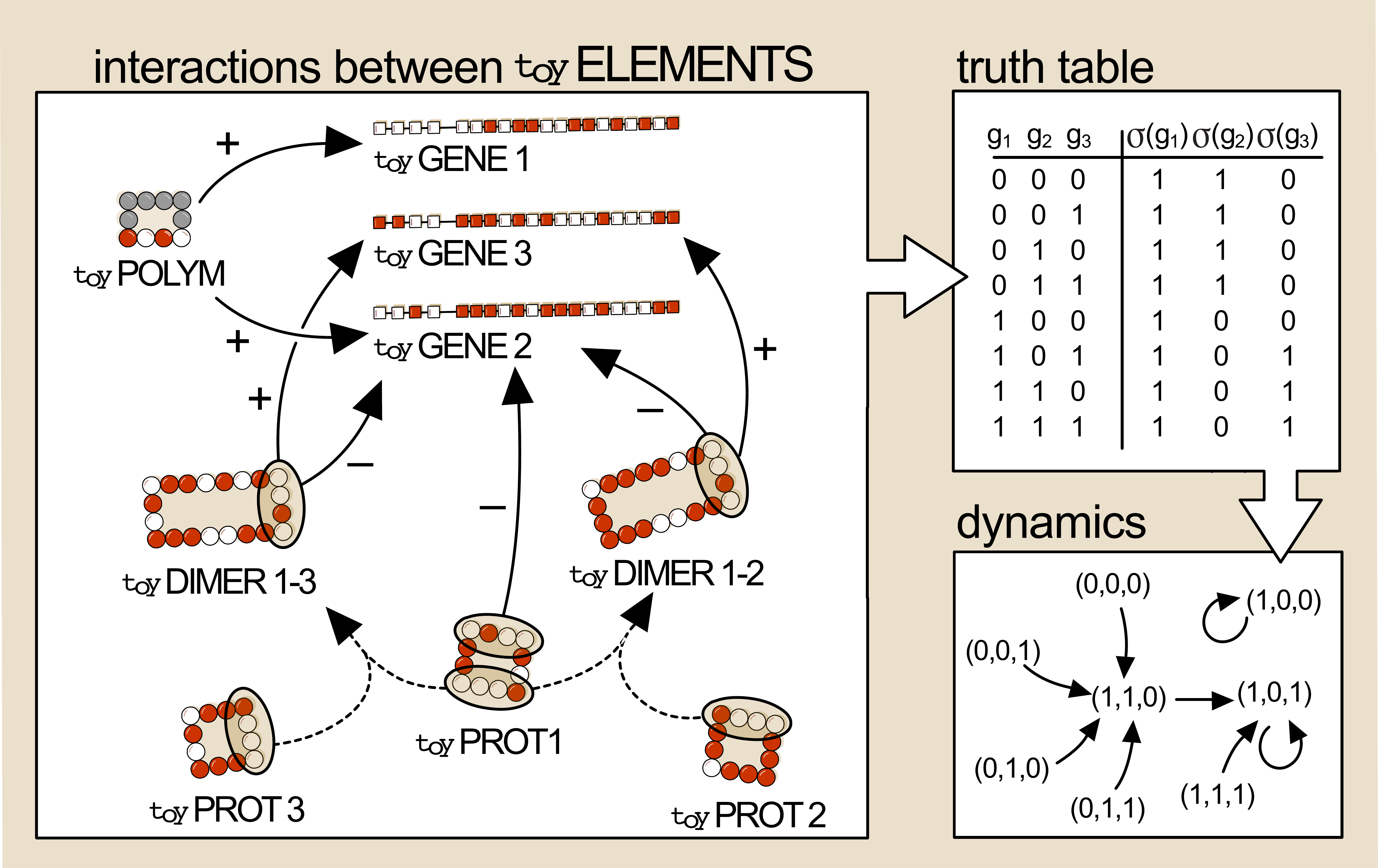}}
\caption[]{\small\textbf{Example of a Boolean network produced by \toyLIFE{} rules.} 
The inputs of the truth table (possible initial states) are all combinations of states of three toyGenes. 
Whenever a toyGene is active, the protein it codes for is
present. The main panel schematically represents all relevant interactions between molecules: in 
this case the toyPolymerase may bind to the promoter regions of
toyGenes 1 and 2 ($+$ signs), and toyProtein 1 inhibits the expression
of toyGene 2 ($-$ signs). The simultaneous presence of toyProteins 1
and 3 leads to toyDimer 1-3, and the simultaneous presence of
toyProteins 1 and 2 to toyDimer 1-2. Both toyDimers inhibit the
expression of toyGene 2 and activate the expression of toyGene 3. The
construction of the Boolean functions codified in the  
truth table is straightforward given the interactions conditional on presence
or absence of each toyProtein (expression or non-expression of each
gene). 
When the truth table
is represented as a directed graph
(summarizing the dynamics of the system from all possible initial
conditions) it is seen that there are two attractors 
for the dynamics: $(1,0,1)$, whose basin of attraction has
size 7, and $(1,0,0)$, whose basin of attraction has size 1. (Note that the
order of toyGenes in a genome is irrelevant, and only responds to
aesthetic reasons.)}
\label{fig:truth}
\end{figure}

If the initial set is formed by $k$ genes, we should consider $2^k$ different
possible vectors of dimension $k$ that correspond to the initial states (i.e.
all combinations of genes being expressed (1) or not expressed (0)). First, the
presence of possible toyDimers coming from expressed genes is evaluated, and
then their interactions with promoter regions (in competition or cooperation
with the toyPolymerase and other toyProteins) are evaluated. This yields an
updated set of expressed toyGenes (a different state) to which the previous
rules are again applied. In this way, one can construct a truth table that can
be subsequently represented in the form of a directed graph (indicating which
state maps into which other) and is fully analogous to a deterministic Boolean
network. An example of a Boolean network derived from a system of three genes
is represented in Figure \ref{fig:truth}.

\subsection*{Boolean networks of \toyLIFE{} depend on metabolism}

The presence of toyMetabolites may modify toyGRN by changing the output states
of the corresponding Boolean network (Figure fig:truthmet). According to the dynamical
rules of \toyLIFE{}, toyMetabolites may interact with toyProteins or toyDimers.
Any molecule bound to a toyMetabolite is no longer available to bind to
promoters, and therefore the expression of the toyGRN is modified. 

toyMetabolites that do not bind to any toyProtein or toyDimer will not be in
the input set for the next time step. On the other hand, bound toyMetabolites
will, unless catalysis occurs. Catalysis happens when, once a bound state
toyMetabolite-toyDimer is formed, an additional toyProtein binds to one of the
units of the toyDimer with an energy lower than that of the initial toyDimer.
In this case, the latter disassembles in favor of the new toyDimer, and in the
process the toyMetabolite is broken (see Figure fig:metabolism for an illustration of the
catalysis process). The two pieces of the toyMetabolite will contribute to the
input set at the next time step. An example of how a toyGRN might change can be
derived from Figure \ref{fig:truth}: if a toyMetabolite able to bind to
toyDimer 1-3 is added to the input set, state $(1,0,1)$ is mapped to $(1,1,0)$
(Figure fig:truthmet). 

\subsubsection*{Metabolons}

The behavior just described prompts the identification of metabolically
functional genotypes that we term metabolons.\footnote{The term metabolon was
first proposed by Paul A. Srere \cite{srere:1985} in 1985 to refer to a
``supramolecular complex of sequential metabolic enzymes and cellular
structural elements'', and is here used as a conceptual analogue.} A metabolon
in \toyLIFE{} is an ensemble of toyGenes able to catalyze at least one
toyMetabolite. In the example above, the three toyGenes are a basic metabolon
that catalyzes in particular the toyMetabolite used as example. 

When the toyMetabolite is absent, the dynamics is described in Figure
\ref{fig:truth} and eventually converges to the steady state $(1,0,1)$
---except if the initial state is $(1,0,0)$. This state is however disturbed
under a constant supply of toyMetabolites able to bind to toyDimer 1-3. In that
case, toyGenes 1 and 2 are expressed in the next time step. toyProtein 1 is
able to form a toyDimer with itself, binding to unit 1 of the
toyDimer-toyMetabolite complex. This latter interaction (which forms toyDimer
1-1) is favored over toyDimer 1-3 and catalysis of the toyMetabolite occurs(see
Figures fig:truthmet and fig:metabolism). If at the next time step the two pieces of toyMetabolite
are unable to interact with any of the toyProteins in the system, they are
eliminated. The toyGRN of this example remains in the new steady state
$(1,1,0)$ ---which is also able to catalyze--- as long as toyMetabolites are
supplied. The three toyGenes system returns to the former steady state
$(1,0,1)$ as soon as the external supply stops.  A graphical summary of a
metabolon in \toyLIFE{} is provided in Figure fig:operon.

\subsection*{The genotype-phenotype map in \toyLIFE}

\toyLIFE{} integrates several levels of complexity: genotypes (sequences of
toyGenes) expressing toyProteins (first level) that interact among themselves
and with promoters generating toyGRNs (second level), and interactions with the
environment (third level) trough catalysis of toyMetabolites. Genotypes are
easily identified as the sequences of Hs and Ps making up toyGenes. In
\toyLIFE, the visible expression of the genotype is best represented through
its interaction with the environment, that is with toyMetabolites. Accordingly,
the phenotype of a genotype (a collection of toyGenes) is defined as the
ensemble of toyMetabolites it can catalyze. There are $2^8=256$ different
toyMetabolites of size 8, and a genotype can either catalyze (in which case it
is a metabolon) or not each of them. The phenotype is formally defined as a
vector of dimension $256$ whose components take value 1 at those positions
corresponding to toyMetabolites that can be catalyzed, and value 0 otherwise.
This definition is analogous to others in the literature where metabolic
activity is explicitely modelled \cite{matias:2009}. 

\subsubsection*{Point mutations might affect different levels}

As it has been defined, there are no mutations of toyGenes explicitly
considered in the dynamics of \toyLIFE. The initial sequences of toyGenes
remain constant as we study properties of the emerging toyGRN and related
phenotypes. This nonetheless, those conditions do not prevent an analysis of
the effect of mutations in the phenotype. Actually, an interesting product of
the multi-level structure of \toyLIFE{} is the possibility of determining at
which level is the effect of point mutations observed. Point mutations are
changes from a P toyN to an H toyN, or {\it vice versa}, in the sequence of a
toyGene.

A summary of changes caused by point mutations in the metabolon in Fig.
\ref{fig:truth} can be found in Table~1. Those effects are not exclusive, that
is, a mutation causing a change in the perimeter of a toyProtein can leave
other functions unchanged, or modify Boolean functions in different ways which
might eventually cause ---or not--- a phenotypic change. Out of the $60$
possible mutations ($12$ affecting the promoters and $48$ affecting the coding
regions), $15\%$ are neutral, $83.3\%$ are deleterious, and only $1.7\%$ are
beneficial. The latter are defined as those mutations enabling the genotype to
catalyze more toyMetabolites than before. Out of $50$ deletereous mutations,
$49$ are lethal (that is, $81.7\%$ of the total). This is a very high
percentage of lethal mutations, compared with an average metabolon ---the
average percentage of lethal mutations, in $10^4$ metabolons chosen at random,
is $56.5 \%$. However, note that this metabolon is not special in any way: in
particular, since it is not a product of evolution and selection, it needs not
have high robustness {\it a priori}. The exploration of genotype space through
neutral paths can likely lead to metabolons with specific properties, as a
higher number of neutral neighbors or a decreased effect of mutations on
phenotype. 

\begin{table}
\begin{center}
\begin{tabular}{l|c|c|c|c|c}
{\bf Effect of mutations} & {\bf Tot} &{\bf Neu} & {\bf Adv} & {\bf
  Del} & {\bf Let}\\
\hline \hline
Different toyProtein folding (same perimeter \& Boolean function) & 3 & 3 & 0 & 0 & 0 \\
Different toyProtein folding \& perimeter (same Boolean function) & 4 & 3 & 0 & 1 & 1 \\
Different toyProtein folding \& perimeter \& Boolean function & 41 & 1 & 1 & 39 & 38\\
Changes in Boolean functions due to the promoter & 12 & 2 & 0 & 10 & 10 \\
\hline
& 60 & 9 & 1 & 50 & 49
\end{tabular}
\end{center}
\caption{\small\textbf{Effect of point mutations in the genotype of
    the example metabolon shown in Fig.~\ref{fig:truth}.} Each of the
  $60$ possible mutations ($12$ affecting the promoters and
  $48$ affecting the coding regions) can have different effects, listed here. For
  each type, the table shows their total number (\textbf{Tot}) and how
  many of them are neutral (\textbf{Neu}), advantageous (\textbf{Adv}), deleterious
  (\textbf{Del} and lethal (\textbf{Let}) mutations. There are a
  total of $9$ neutral mutations  ($15\%$), $1$ advantageous mutation
  ($1.7\%$), and $50$ deleterious mutations ($83.3\%$). Of the latter,
 $49$ ($81.7\%$ of the total) are lethal mutations.}
\label{tab:mutations}
\end{table}

\subsubsection*{Functional properties of three-toyGenes genotypes}

The genotype-phenotype map in \toyLIFE{} is highly redundant and displays ample
variations in the number of genotypes representing the same phenotype.
Redundancy comes not only from neutral mutations, but also from the existence
of compensatory mutations and genomic solutions with mutations in many toyN
that yield the same phenotype. The redundancy of the HP model has been
discussed in the literature \cite{li:1996,holzgrafe:2011} and is, through the
interaction rules of \toyLIFE{}, non-trivially extended to the formation of
molecular aggregates and catalytic processes. These are qualitative features
that \toyLIFE{} shares with natural systems and that we quantify in the
following. 

\begin{figure}[!b] 
\includegraphics[angle=270, width=150mm]{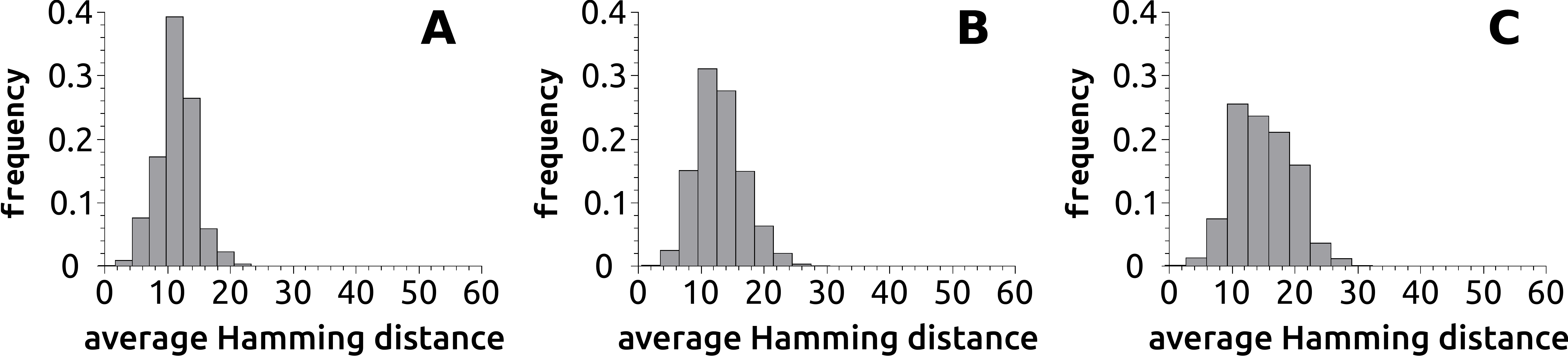}
\caption[]{\small\textbf{Histograms of Hamming distances from an
    ancestral genotype obtained through neutral paths.} We chose
  a sample of $10^4$ three-toyGene genotypes at random and, for each of them,
  computed $10^4$ neutral random walks (see text). For each random walk, we then measured
  the Hamming distance between the final genotype and the original one, 
  and averaged this distance over all random walks. The
  histograms show the distribution of average Hamming distances for
  all $10^4$ genotypes. We repeated
  this experiment with random walks of length $10^2$ (\textbf{A}),
  $10^3$ (\textbf{B}) and $10^4$ (\textbf{C}). The average distance
  grows with the length of the random walks: from $12.4$ (\textbf{A})
  to $14.0$ (\textbf{B}) to $16.6$ (\textbf{C}). Note that the width
  of the distributions also grows with length of the random walk.}
\label{fig:HammingD}
\end{figure}

We begin by analyzing the navigability of the genotype space. To this end we
use metabolons similar to the one represented in Fig. \ref{fig:truth} and
perform random walks on their neutral space. That is, we take three initial
gene sequences, which form a genotype (or genome) of length $60$. After making
sure this genome is able to catalyze at least one toyMetabolite, we attempt a
point mutation at a randomly chosen genome site. If the phenotype of the mutant
is identical to that of the previous genome, the mutation is accepted;
otherwise, the mutation is discarded and, in either case, the process is
repeated. Mutations do not affect the toyPolymerase. The mutation process is
attempted a variable number of times (that is, the random walks are of
different lengths: $10^2$, $10^3$ or $10^4$), and repeated for a large number
of independent realizations ($10^4$ random walks for each one of $10^4$
original genotypes). In this way, we obtain the histograms shown in Figure
\ref{fig:HammingD}. The average number of accumulated substitutions, i.e. the
Hamming distance between the original genome and the current one, grows with
the number of mutations attempted, yielding genomes that increasingly differ
from their ancestors. This behavior is fully analogous to that observed in RNA
secondary structure neutral networks \cite{schuster:1994}, in proteins
\cite{babajide:1997}, and in one-level models of gene regulatory networks
\cite{ciliberti:2007} or metabolism \cite{matias:2009}. 

\begin{figure}[!b] 
\includegraphics[angle=270, width=150mm]{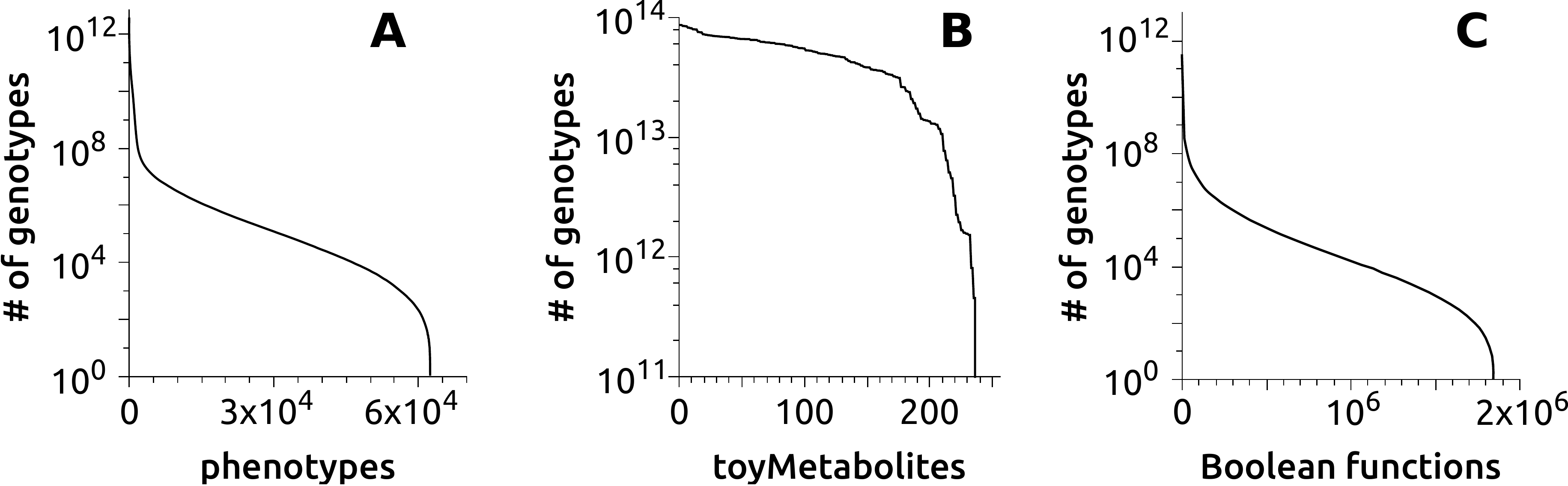}
\caption[]{\small\textbf{Statistical properties of three-toyGene genotypes.} \textbf{A:} Phenotype abundance. There are $66,720$ different phenotypes with 
abundances that vary in eleven orders of magnitude. The phenotypes, in
the $x-$axis, are
rank ordered following the number of genotypes that
express them. \textbf{B:} Number of metabolons able to break a given toyMetabolite. The latter, in the $x-$axis, are 
rank-ordered following the number of genotypes that catalyze
them. \textbf{C:} Abundances of Boolean functions in phenotypes. Some Boolean functions are easy to 
obtain, while others are very rare. Again, Boolean functions are rank
ordered according to the number of genotypes that express them.}
\label{fig:operon_statistics}
\end{figure}

Next, we have exhaustively explored the space of genotypes consisting of three
toyGenes and evaluated their ability to break toyMetabolites of size 8. In
total, there are around $1.4 \times 10^{14}$ metabolons out of the total of
$\sim 2 \times 10^{17}$ three-toyGenes genomes ---the number of combinations of
all possible toyGenes, $2^{20}$, in groups of three, with repetitions. That is,
only about $0.07\%$ of all possible genomes are able to catalyze toyMetabolites
of size 8. In agreement with the definition of phenotype given above, there are
up to $2^{256} \approx 10^{77}$ different phenotypes.  However, only $66,720$
different phenotypes can be realized by three-toyGene genomes, yielding an
average close to $2 \times 10^9$ metabolons per phenotype. This average is
however not very informative, since the variation in phenotype abundance is
enormous (Fig. \ref{fig:operon_statistics}A). There is also ample variability
in the characteristics of phenotypes. Most toyMetabolites are broken by around
$10^{14}$ genomes, but some of them can be broken by far fewer genomes (see
Fig. \ref{fig:operon_statistics}B), and some toyMetabolites cannot be broken by
any genome at all. Specifically, there are $20$ toyMetabolites that cannot be
broken. They have a particular composition or structure, since they contain $7$
consecutive H or P sugars (there are $4$ such toyMetabolites) or are
palindromes (a total of $16$ additional toyMetabolites). In both cases, only
symmetrical toyDimers can bind to these toyMetabolites ---asymmetrical
toyDimers give rise to ambiguous interactions and are discarded. But
symmetrical toyDimers bound to a given toyMetabolite cannot be broken by any
toyProtein, because both subunits forming the toyDimer have the same perimeter
and, again, this gives rise to ambiguous interactions.

Finally, many Boolean functions are obtained from different genotypes. For $n$
genes, there are ${(2^n)}^{2^n}$ different Boolean functions, because for each
of the $2^n$ possible inputs there are $2^n$ possible outputs. For three genes,
this is already a very large number, $8^8 = 16,777,216$ Boolean functions.
Figure \ref{fig:operon_statistics}C represents the abundances of Boolean
functions. As can be seen, there is a highly unequal representation in terms of
genotypes, and only about 10\% of all possible Boolean functions are actually
represented by at least one genotype. 

\section*{Discussion}

Despite their simplicity, models of the genotype-phenotype map provide
important conceptual insights. Not only that, some of them have been able to
capture qualitative and quantitative features of the natural systems they aimed
at representing. However important details might be, these are occasionally
offset by universal rules that determine the emerging phenomenology and
statistical behavior both of biological systems and their {\it in silico}
cartoons. For instance, the HP model of protein folding, which disregards the
fine chemical structure of aminoacids and constrains HP polymers to fold on
regular lattices, is able to predict the existence of unique folding states for
sufficiently large polymers and the formation of hydrophobic cores, among
others, in agreement with empirical knowledge \cite{lau:1989}. Computational
studies of RNA sequences folding into their minimal energy secondary structure
have enlightened a large number of dynamical and structural properties with a
clear empirical counterpart, such as punctuated equilibria at the molecular
level \cite{huynen:1996} or increases in robustness with phenotype size
\cite{aguirre:2011}, a feature that is quantitatively shared by all
genotype-phenotype maps studied to date
\cite{li:1996,bloom:2007,greenbury:2014,dallOlio:2014}. Boolean networks,
despite working with a sharp threshold for gene expression, have witnessed
notable success, including faithful reproduction of living cell cycles
\cite{davidich:2008}. \toyLIFE{} constructs a multi-level genotype-phenotype
map from simple interactions inspired by the HP model from which the logical
architecture of Boolean networks emerges. The addition of metabolic abilities
arises as a natural extension of the basic model. 

The possibilities of \toyLIFE{} are not exhausted by the cases presented in
this work, which constitute a minimal ---hopefully illustrative--- sample of
the kind of complexity \toyLIFE{} might encode for. In devising the model here
analyzed, we had to make some choices regarding energy parameters, number of
molecules or genes allowed to interact, or disambiguation rules to define
functional molecules. Preliminary analyses of equally reasonable alternatives
indicate that \toyLIFE{} universes defined through similar rules display a
phenomenology comparable to the one here presented. Still, a deeper exploration
of certain emergent behaviors seems worth pursuing. First, \toyLIFE{} gives
clues on the level ---between genotype and phenotype--- where the effect of
mutations can be seen. Distance between phenotypes is simple to define in
\toyLIFE, and a more systematic analysis might allow as well a quantitative
comparison with empirical studies measuring the distribution of fitness effects
\cite{eyrewalker:2007}. This function is an important object in developing
models of phenotypic change that effectively incorporate the molecular details
of evolution. Second, even the three-toyGenes genomes here studied reveal the
emergence of functional abilities not implemented in the basic rules of the
model, such as toyProteins behaving as conditional activators. This observation
indicates that a protein can be recruited in appropriate molecular contexts to
perform additional functions, that is, it can be co-opted to develop a second
useful, but non-adaptive, role \cite{piatigorsky:2007}. The consideration of
larger genomes and larger molecular aggregates should certainly usher in new
collective abilities, and very often lead to multi-functional toyProteins. In
this scenario, the effect of single mutations might then arise at multiple
levels, likely revealing a pleiotropic structure \cite{gwagner:2011} in the
\toyLIFE{} genotype-phenotype map. The effect of point mutations at different
levels and the fraction of neutral or lethal mutations, among others, would be
relevant issues to explore. It will also be interesting to study how different
metabolons are fit together in a larger genome, developing more complex
metabolic networks than the ones shown in this paper.  Third, it would be worth
comparing the statistical properties of random Boolean networks and other gene
regulatory networks with those obtained from \toyLIFE. An open and challenging
question is how an explicit consideration of genome dynamics modifies or
constrains the statistical properties of genotype-phenotype models that discard
them. Fourth, for three-toyGenes genomes, we have observed a very high dilution
of metabolons in comparison to genomes that cannot break any of the
toyMetabolites considered. This result is in qualitative agreement with models
of gene regulatory networks \cite{ciliberti:2007} and metabolism
\cite{matias:2009} that ignore lower levels. An open question is how this
dilution changes as we increase the number of participating toyGenes and
diversify the set of toyMetabolites that should be catalysed. At present, this
study is severely limited by the computational time it requires. Finally, it is
easy to implement additional mutational mechanisms in \toyLIFE, such as gene
duplication or deletion. The implications of such a change on the phenotype
cannot be foreseen without an explicit analysis. However, we have found
two-toyGenes metabolons whose function is maintained when a third toyGene is
added. In this respect, \toyLIFE{} might provide complementary insight on the
evolutionary effects of gene duplication, including their lethality and their
ability to develop new functions \cite{lynch:2000,kondrashov:2012}.

\section*{Conclusions}

The design of models able to shed light on the complex structure of the
genotype-phenotype map is actively pursued. Research in the field has led to
the identification of a number of generalities, as the broadly varying
robustness of genotypes, the unequal distribution of phenotype sizes or the
existence of genotype networks that permit the navigability of the space of
genotypes ---and thus adaptation and evolutionary innovation--- while
preserving function. However, some other issues still represent a challenge, as
the effect of mutations in fitness or the precise mechanisms constraining
evolution as a result of the bottom-up and top-down coupling among intermediate
levels in the genotype-phenotype map. \toyLIFE{} and possible variants thereof
might provide a useful framework to explore these and related questions. 


\section*{Methods}

\subsection*{Folding of toyProteins and molecular interactions in \toyLIFE}

The folding of toyProteins exactly follows the folding of HP random polymers
first introduced in \cite{dill:1985}. We consider sequences of length 16 and
limit the possible folds to compact $4 \times 4$ structures on a lattice. There
are 38 such structures ignoring symmetries. The energy of a fold is the sum of
all pairwise interaction energies between toyAminoacids that are not contiguous
along the sequence. Pairwise interaction energies are $E_{\text{HH}}=-2$,
$E_{\text{HP}}=-0.3$ and $E_{\text{PP}}=0$, as in \cite{li:1996}. The structure
of a toyProtein is its lowest energy fold. If there is more than one fold with
the same minimum energy, we select the one with fewer H toyAminoacids in the
perimeter. If still there is more than one fold fulfilling both conditions, we
discard that protein by assuming that it is intrinsically disordered and thus
non-functional \cite{radivojac:2007}. Out of $2^{16} = 65,536$ possible
proteins, $37,525$ do not yield unique folds.  We find $6,290$ different
toyProteins with $568$ different perimeters.

toyProteins interact through any of their sides with other toyProteins (to form
toyDimers), with promoters of toyGenes, and with toyMetabolites. Once formed,
toyDimers can also bind to promoters or toyMetabolites through any of their
sides (binding to other toyProteins or toyDimers is not permitted). In all
cases, the interaction energy ($E_{\text{int}}$) is the sum of pairwise
interactions for all HH, HP and PP pairs formed in the contact. Bonds can be
created only if the interaction energy between the two molecules
$E_{\text{int}}$ is lower than a threshold energy $E_{\text{thr}}=-2.6$. Note
that a minimum binding energy threshold is necessary to avoid the systematic
interaction of any two molecules. Other alternatives might be the addition of
terms that represent an energetic cost\footnote{In other models, as in RNA
folding, the threshold is set to 0 because structural elements such as loops or
dangling ends yield positive contributions to the total folding energy.} or the
consideration of stochastic interactions, such that those with higher energy
would be less probable. If below threshold, the total energy of the resulting
complex is the sum of $E_{\text{int}}$ plus the folding energy of all
toyProteins involved. The lower the total energy, the more stable the complex.
When several toyProteins or toyDimers can bind to the same molecule, only the
most stable complex is formed. Consistently with the assumptions for protein
folding, when this rule does not determine univocally the result, no binding is
produced. 

toyPolymerase is a special kind of toyProtein. It only has one interacting side
(with sequence HPHP) and its folding energy is fixed to value $-15$.
toyPolymerase binds to promoters or to the right side of a toyProtein/toyDimer
already bound to a promoter. In either case the toyGene is expressed.

As the length of toyMetabolites is usually longer than 4 toyN (the length of
interacting toyProteins sites), there might be several positions where the
interaction with a toyProtein has the same energy. In those cases we select the
sites that yield the most centered interaction. If ambiguity persists, no bond
is formed. Also, no more than one toyProtein/toyDimer is allowed to bind to the
same toyMetabolite, even if its length would permit it. toyProteins/toyDimers
bound to toyMetabolites cannot bind to promoters.

Binding to promoters is decided in sequence. Starting with any of them (the
order is irrelevant), it is checked whether any of the toyProteins/toyDimers
available bind to the promoter, and then whether toyPolymerase can subsequently
bind to the complex and express the accompanying coding region. If it does, the
toyGene is marked as active and the toyProtein/toyDimer is released. Then a
second promoter is chosen and the process repeated, until all promoters have
been evaluated.  toyGenes are only expressed after all of them have been marked
as either active or inactive. Each expressed toyGene produces one single
toyProtein molecule. There can be more units of the same toyProtein, but only
if multiple copies of the same toyGene are present.

\section*{Acknowledgments}
This work was supported through projects FIS2011-­22449 (CFA, PC and
JAC) and FIS2011--27569 (SM) of the Spanish MINECO.


\newpage

\section*{Supporting Information}

\begin{figure}[!t]
\centerline{\includegraphics[width=150mm]{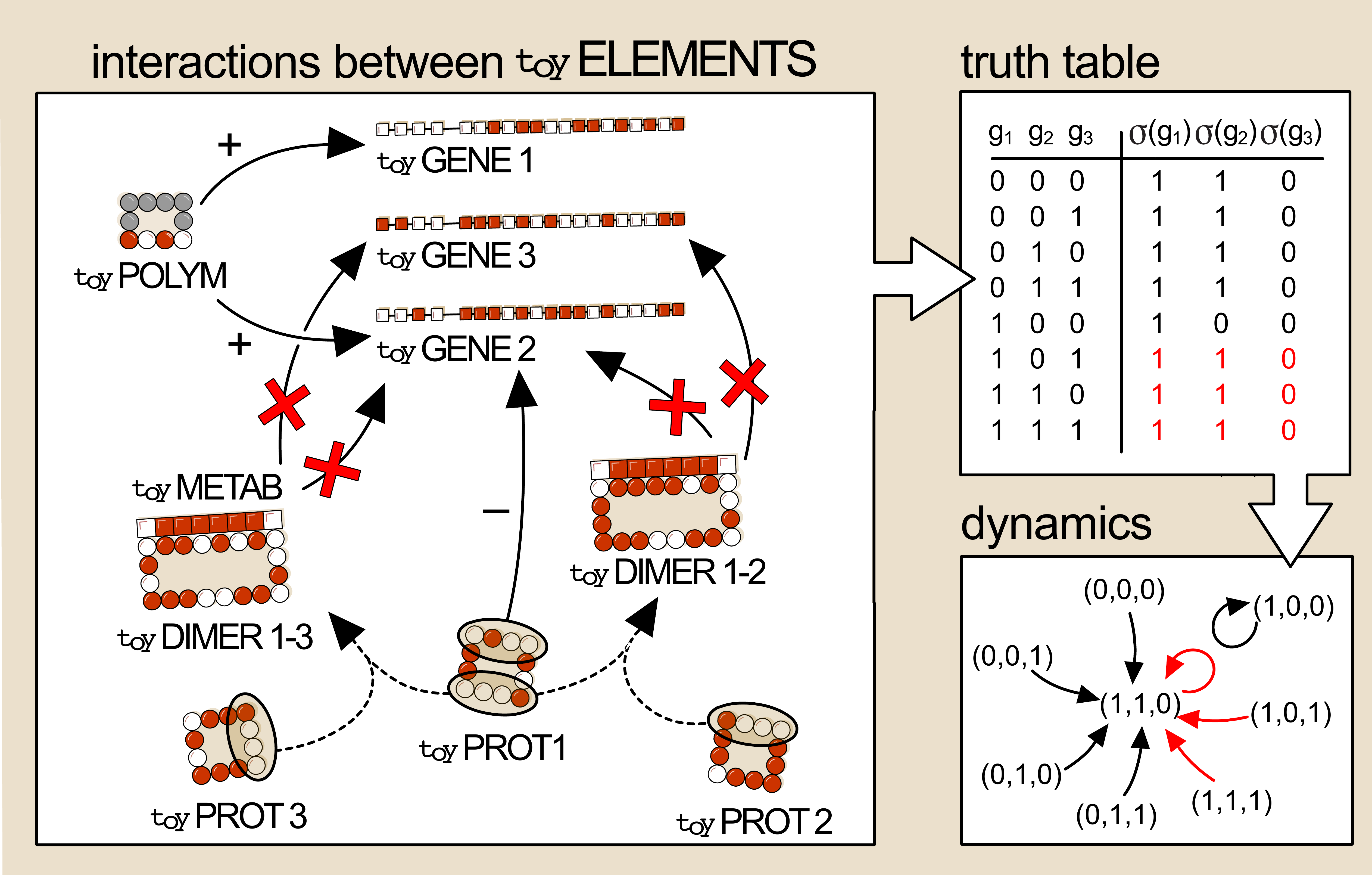}}
\caption[]{\small\textbf{toyMetabolites change the expression of
    toyGRNs.} This is the same example illustrated in Figure~3 (main text), but
  with the addition of a toyMetabolite able to bind toyDimers 1-2 and 1-3. When these toyDimers
bind to a toyMetabolite, they no longer participate in the regulation
phase, and thus states $(1,0,1)$, $(1,1,0)$ and $(1,1,1)$ are all
mapped to state $(1,1,0)$ in the
presence of a toyMetabolite. In other words, the presence of the
toyMetabolite changes three entries in the truth table, and therefore the associated
Boolean network ---one of the asymptotic states is now different.}
\label{fig:truthmet}
\end{figure}

\begin{figure}[!b] 
\centerline{\includegraphics[width=100 mm]{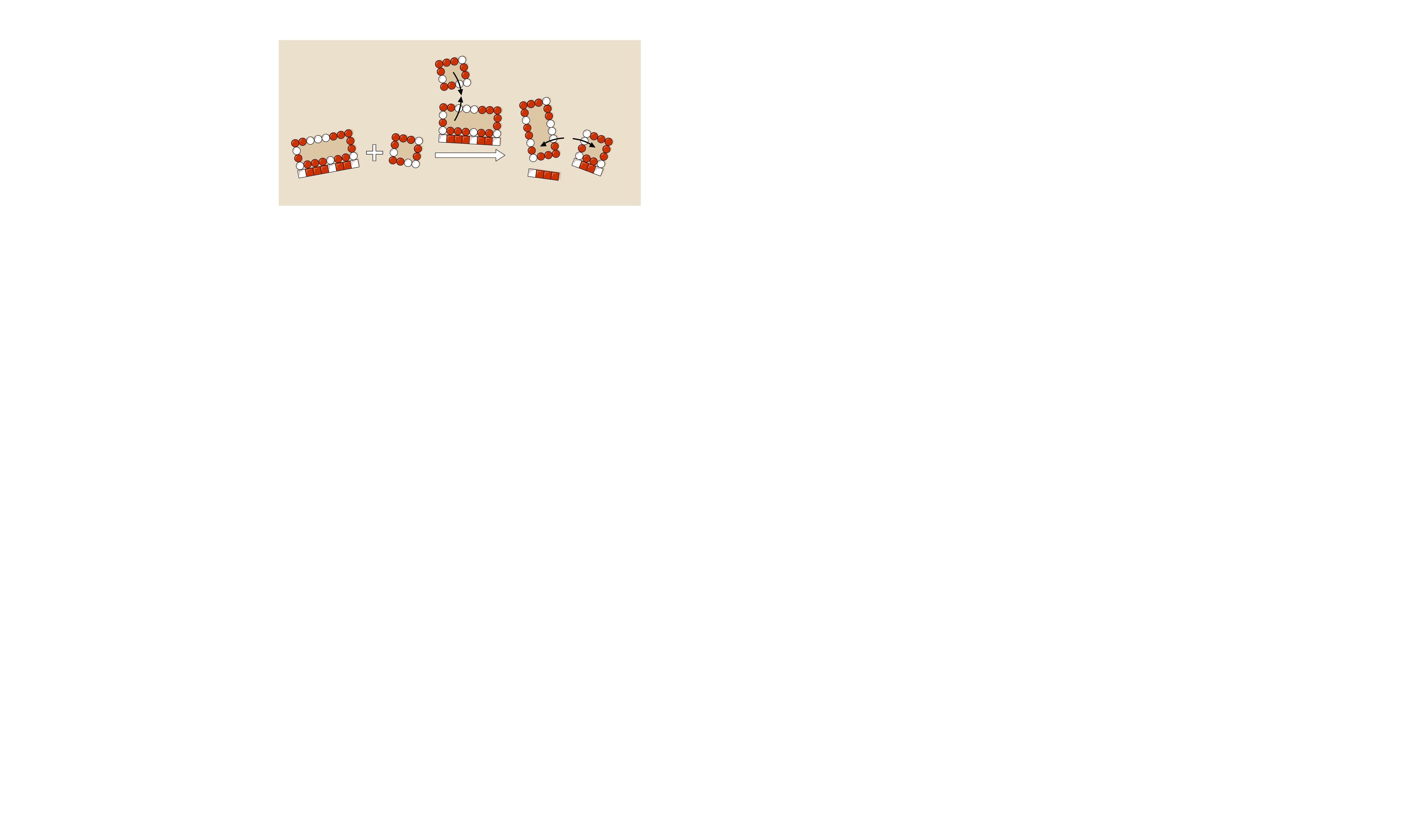}}
\caption[]{\small\textbf{Catalysis in \toyLIFE.} A toyDimer is bound to a toyMetabolite when a new toyProtein comes in. If the new toyProtein binds to
one of the units of the toyDimer, forming a new toyDimer energetically more stable than the old one, the two toyProteins will unbind and
break the toyMetabolite up into two pieces. We say that the toyMetabolite has been catalyzed.}
\label{fig:metabolism}
\end{figure}

\begin{figure}[!t] 
\centerline{\includegraphics[width=150mm]{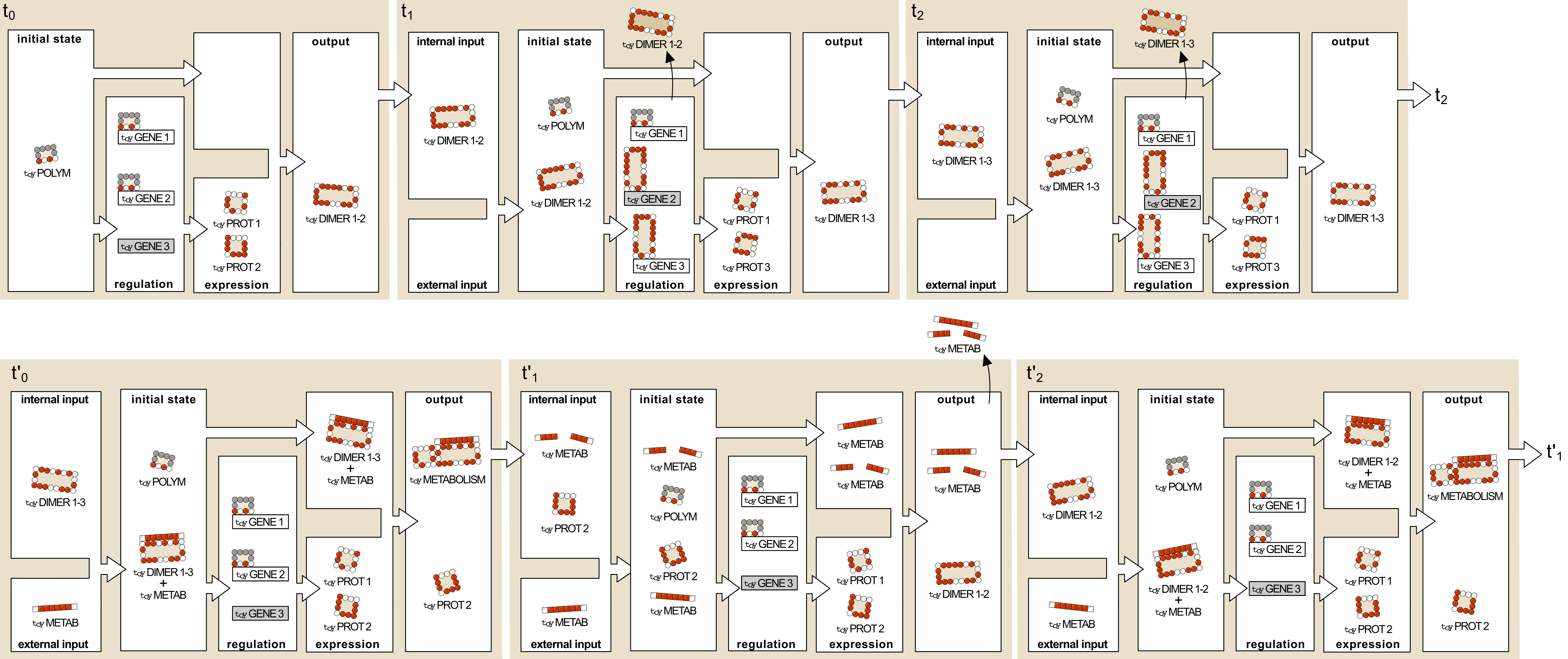}}
\caption[]{\small\textbf{Summary of a metabolon activity in \toyLIFE.}
  Consider the toyGRN of Figures~3 (main text) and \ref{fig:truthmet} (above). Initially ($t_0$)
  all three toyGenes are off.  The toyPolymerase can bind to the
  promoter regions of toyGenes 1 and 2, expressing toyProteins 1 and
  2, and toyDimer 1-2 forms. Thus,  the internal input set for time
  step $t_1$ contains 
  toyDimer 1-2. At the regulation phase in $t_1$ the toyPolymerase
  (which is always present) activates the expression of toyGene 1, and
  toyDimer 1-2
  inhibits the expression of 
  toyGene 2 and activates that of toyGene 3. As
  a result, toyDimer 1-3 forms. The input set for time step $t_2$
  then contains just toyDimer 
  1-3. At $t_2$ toyDimer 1-3 again inhibits the expression of
  toyGene 2 and activates that of toyGene 3, and the internal input
  set for next time step will again only contain toyDimer 1-3. The toyGRN has
  reached a  steady state. But if at this point a toyMetabolite is
  added to the input set, the behavior of the toyGRN changes
  (below). The toyMetabolite is such that it binds toyDimer  1-3, so the
  toyDimer is unable to participate in regulation, and the
  toyPolymerase activates the expresion of toyGenes 1 and
  2. toyProtein 1 is then able to bind to toyDimer 1-3 in the output
  phase, breaking it. The internal input set for time step $t'_1$ is
  formed by toyProtein 2 and the rests of the broken
  toyMetabolite. Even if the toyMetabolite appears again as a external
  output, nothing can bind it in the input phase, so this does not
  affect regulation. toyProtein 2 has no effect on regulation, and
  again toyProteins 1 and 2 are expressed, and toyDimer 1-2 is
  formed. As nothing has bound the toyMetabolites, they will not be
  present in the internal input set of time step $t'_2$, which will
  only contain toyDimer 1-2. If a new toyMetabolite is provided as
  the external output, the toyDimer will bind to it, and the cycle
  begins again (however, note that from now on all metabolism will be
  due to toyDimer 1-2 instead of toyDimer 1-3).}
\label{fig:operon}
\end{figure}

\end{document}